\documentclass[11pt,twoside]{article}

\usepackage{asp2006}
\usepackage{epsf}
\usepackage{psfig}
\usepackage{lscape}

\begin{document}
\title{Starbursts in Isolated Galaxies: the influence of the ISM model}   %%% Fill in title
\author{Christian Theis}   %%% Fill in author names
\affil{Institute of Astronomy, University of Vienna, T\"urkenschanzstr.\ 17, 1180 Vienna, Austria}   %%% Fill in title
\author{Joachim K\"oppen}   %%% Fill in author names
\affil{Observatoire Astronomique de Strasbourg, 11, rue de l'Universit\'e, F-67000 Strasbourg, France}    %%% Fill in author affiliations

\begin{abstract} %%% Abstract to run on from here.

  We study the stability properties of isolated star forming dwarf
  galaxies which undergo dynamically driven star bursts induced by
  stellar feedback.  Here we focus on the impact of the adopted ISM
  model, i.e.\ either a diffuse or a clumpy ISM. We apply a one-zone
  model extended for active dynamical evolution.  This allows for a
  coupling between the dynamical state of the galaxy and its internal
  properties like star formation activity or the thermal state (or
  dynamical pressure, respectively) of the interstellar medium (ISM).

We found two major types of repetitive star bursts: one set (type A)
of quasi-periodic star bursts is related to the dynamical timescale of
the galaxy. In that case, the star formation follows the variations of
the gas density induced by decaying virial oscillations. The second
set (type B) of recurrent star bursts is characterized by a long
quiescence period given by the sum of the dynamical and the
dissiptional timescale: after a first burst, the inserted energy leads
to a substantial expansion of the system, by this stopping any
significant SF activity. A next burst might occur, when the gas
reaches high densities again, i.e.\ after the gas recollapsed and the
energy injected by stellar feedback is dissipated.

In case of a diffuse ISM model, type A bursts are the most common type
due to the high efficiency of radiative cooling (no type B bursts are
found).  Bursts occur then mainly during an initial transitory phase.
In case of a clumpy ISM model (i.e.\ dissipation by inelastic
cloud-cloud collisions), the dissipative timescale is of the order of
the dynamical time or longer. This allows for both, type A and type B
bursts.  Whereas initial transitory bursts are quite common, type B
bursts are only found in a small mass range for given feedback and
dissipation parameters.

\end{abstract}

%%% MAIN BODY OF TEXT GOES HERE. CONSULT "INSTRUCTIONS FOR AUTHORS USING
%%% LATEX2E MARKUP", SECTIONS 2.3-2.6 FOR HELP WITH EQUATIONS, FIGURES,
%%% AND TABLES.

%\section{}   %%% Top level section head (remove "%" symbol)
%\subsection{}   %%% Second level section head (remove "%" symbol)
%\subsubsection{}   %%% Lowest level section head (remove "%" symbol)
%\section*{}    %%% Unnumbered top level section head (remove "%" symbol)
%\subsection*{}   %%% Unnumbered second level section head (remove "%" symbol)

\section{Introduction}

Starbursts are a well known phenomenon in some dIrrs (e.g.\ Gallagher \&
Hunter \cite{gallagher84}; for a critical discussion of the exact 
definition of starbursts see Knapen \& James \cite{knapen09}). They imply a mechanism to organise
large-scale star formation which cannot operate all the time due to
fast gas consumption.  Moreover, the instability leading to an
episodic SFR should work in a small parameter range only, because
most of the isolated dIrrs seem to evolve in a highly self-regulated
manner.

The first theoretical models dealing with the variability of the star 
formation rate in
galaxies were based on closed-box models. For reasonable models of the
interstellar medium (ISM) it turned out that stellar feedback is rather
efficient in suppressing instabilities (e.g.\ Ikeuchi \&
Tomita \cite{ikeuchi83}, Scalo \& Struck-Marcell \cite{scalo86}).
These models only allow for a burst-like behaviour when a long
time-delay between star formation and stellar feedback is introduced.
K\"oppen et al.\ \citep[][hereafter KTH95]{koeppen95} corroborated
this result by investigating a simple numerical model motivated by
full chemodynamical simulations.  They demonstrated that the effective
SFR is almost independent of the detailed recipe for the stellar birth
function, provided a negative feedback due to the thermal state of the
ISM is considered in the stellar birth function.

These box models, however, suffer
from neglecting galactic dynamics.
Several attempts have been made to study the evolution of isolated
dwarf galaxies by taking full dynamics, as well as
star formation and stellar feedback, into account 
(Kruegel et al.\ \cite{kruegel83}, and later Hensler et al.\
\cite{hensler04}, Pelupessy et al.\ \cite{pelupessy04}, Struck
\cite{struck05}).  Though these calculations differ in many details,
they agree in producing large-scale star formation variations on
dynamical timescales which indicates the
crucial role dynamics plays for episodic starbursts.

A major disadvantage of numerical simulations is their complexity.
This often prevents a detailed investigation of the parameter space.
Instead of analysing dwarf galaxies by such detailed models, we apply
a set of equations that is as simple as possible, but still
incorporating the generic features of the complex numerical
simulations. We use an extended version of the analytical models of
KTH95 that allows us to deal with the dynamical evolution of a galaxy
in a simple way (for details cf.\ to Theis \& K\"oppen
\cite{theis09}). Here we compare the evolution of dwarf
galaxies for two different ISM models: in the diffuse ISM model energy
is dissipated by radiation, whereas in the clumpy ISM model energy is
lost via inelastic cloud collisions.

\section{Results}

\begin{figure}[!ht]
  \plotfiddle{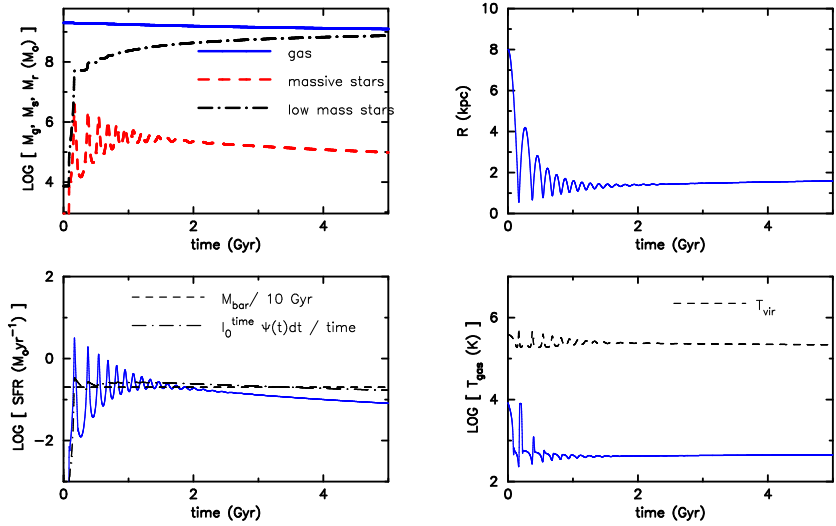}{6cm}{270}{40}{40}{-185}{210}
%  \plotfiddle{mod004b_5gyr_all.eps}{6cm}{270}{40}{40}{-185}{210}
%  \plotfiddle{mod004b_5gyr_all.eps}{8cm}{270}{46}{46}{-205}{245}
  \caption{Evolution of a model with a baryonic mass
    of $2 \cdot 10^9 \, {\rm M}_\odot$. Shown are the masses of the
    baryonic components (upper left), 
%    i.e.\ gas (solid), low-mass
%    stars, and stellar remnants (dot-dashed), and massive stars
%    (dashed), 
    the mean radius of the system (upper right), the total
    SFR (lower left), and the gas temperatures, i.e.\ the actual
    temperature and the virial temperature (lower
    right).  
}
%The SFR is compared with the gas
%    consumption averaged over 10 Gyrs and the mean star formation up to
%    time $t$, i.e.\ $\int_0^t \Psi(t') dt' / t$.  The temperature is
%    compared with the virial temperature of the system including the
%    self-gravity of the baryons and the dark matter contribution.}
  \label{theisfig1}
\end{figure}

\noindent
{\bf Diffuse ISM models.}
Fig.\ \ref{theisfig1} shows the evolution for the diffuse ISM model.
Due to strong cooling, the gas temperature drops almost immediately to
values near $10^4 {\rm K}$ bringing the system far out of virial
equilibrium.  In the ongoing collapse the density increases and the gas temperature drops
even more. The cooling and density enhancement continue, until star formation
becomes more prominent after about 100 Myr. The first stars reinject
energy to the ISM leading to the quasi-equilibrium stage known from
the box models of KTH95.  However, different to them, the density
still increases due to the ongoing collapse.  Though the star
formation and, by this, the stellar energy injection grows, the
feedback cannot stop further collapse, because cooling is too
efficient to prevent the gas from reaching virial equilibrium. The
collapse proceeds until the angular momentum barrier becomes dominant
and the system expands.  Then the first star burst episode finishes
and the gas is quickly heated to $10^4 {\rm K}$. However, due to the strong enhancement
of the cooling rate at temperatures beyond $10^4 {\rm K}$, the
gas never reaches the virial temperature.

The longterm evolution of this model shows that the previously
discussed bursts are just transitory phenomena. The oscillations are
almost completely damped after 2 Gyr.  Such a behaviour, i.e.\ only an
initial burst, was found in almost all models with radiative cooling
as source of dissipation.

%\section{The clumpy ISM model}

\begin{figure}[!ht]
  \plotfiddle{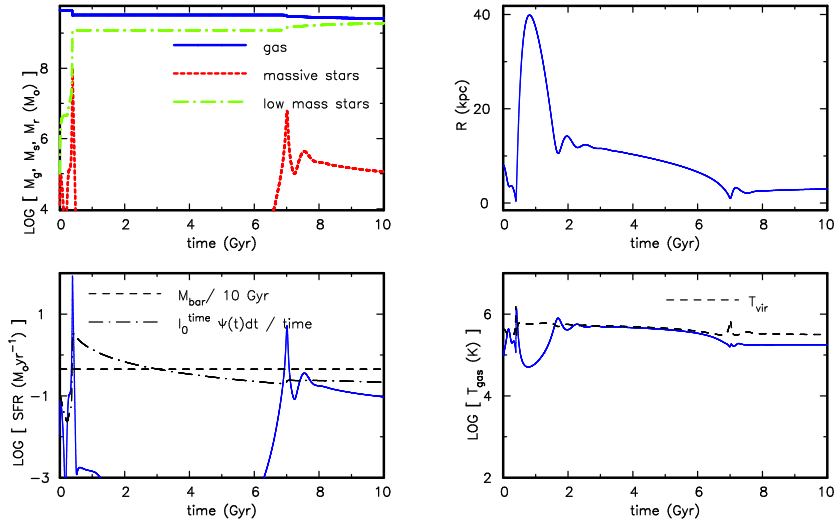}{6cm}{270}{40}{40}{-185}{210}
%  \plotfiddle{mod032_10gyr_all.eps}{8cm}{270}{46}{46}{-205}{245}
  \caption{Temporal evolution of a model with energy dissipation by inelastic
   cloud collisions. The baryonic mass amount to
   $4.5 \cdot 10^9 \, {\rm M}_\odot$ (for details see Fig.\ \ref{theisfig1}). 
%Note that the ''gas temperature'' denotes
%   here the velocity dispersion of the clump system.
}
  \label{theisfig2}
\end{figure}

\noindent
{\bf Clumpy ISM models.}
   The evolutionary behaviour changes when applying the clumpy ISM
model. Though the structure of the equations remains identical, the
dissipational timescale differs substantially from the previous case
in which the cooling timescale was much shorter than other involved
timescales. For cloud-cloud collisions,
the timescale $\tau_{\rm coll}$ becomes of the order of Gyrs.
%\begin{equation}
%   \tau_{\rm coll} \approx 0.97 \left( \frac{M_g}{10^9 M_\odot} \right)^{-1}
%    \left( \frac{R}{5\,{\rm kpc}} \right)^{7/2}
%    \left( \frac{M_{\rm DM}(R)}{10^{10} M_\odot} \right)^{-1/2} \, {\rm Gyr}
% \,\,.
%   \label{taucc}
%\end{equation}

  Driven by stellar feedback and the much longer dissipational time
the ISM expands after the first burst (Fig.\ \ref{theisfig2}, upper
right). The dark matter halo, however, prevents the gas from being
blown-away. Therefore, it collapses again reaching a dynamical
equilibrium after about 2 Gyrs. This can be seen from a comparison between
the actual kinetic temperature (velocity dispersion of the cloud system)
and the virial temperature of the system. The
ongoing evolution is governed by a slow loss of energy due to
dissipation. During this stage of slow shrinkage the evolution is
close to virial equilibrium.  After 6 Gyrs the system lost sufficient
energy to start with the next run-away collapse resulting in another
starburst event after a long period of quiescence. Finally, the
galaxy produces stars on a mean rate of about $0.1 M_{\sun} \mathrm{yr}^{-1}$.

\subsection*{Discussion}
\label{sect_discussion}

  The main difference between the two model scenarios is the ratio $r$
between the dissipational timescale $\tau_{\rm diss}$ and the
dynamical timescale $\tau_{\rm dyn}$. 
In case of the diffuse ISM model (small $r$) the energy
input by the stars is almost instantaneously radiated away. Therefore,
the radial oscillations are only decaying initial virial oscillations
acting on a dynamical timescale.  This behaviour turned out to be
rather robust against variations of model parameters like stellar
heating, gas mass, initial size or stellar birth function.

On the other hand, when $r$ becomes of the order of one or larger
(i.e.\ for the clumpy ISM model), two timescales are important: the
dynamical timescale and the dissipational timescale. After a first
burst, a long period ($\tau_{\rm dyn} + \tau_{\rm diss}$) of
quiescence follows, before the next star formation episode starts.

During most of the quiescence phase the gas is semi-adiabatically accreted
with an average infall velocity of only a few km\,s$^{-1}$.  The related extended
gas reservoirs are known around many dwarf galaxies (e.g.\ NGC 4449).
They are very sensitive to perturbations by other galaxies. E.g.\ a
substantial amount of the gas might be stripped off, by this reducing
the strength of the second burst.

  After the second burst the galaxy enters a phase with continuous star
formation up to the present (here: 10 Gyr). This is in agreement with
observations of several dIrrs which still show a substantial, more or
less continous star formation ongoing over the last one or few Gyrs
(e.g.\ Greggio et al.\ \cite{greggio93}).

%   The existence and duration of a quiescence phase depends on the
%strength of the feedback and the depth of the dark matter potential.
%For a given DM distribution, feedback coefficient $h$ and SBF
%$\Psi_b$, the quiescence period $T_{\rm quiet}$ sensitively depends on
%the baryonic mass: within a factor of 3 $T_{\rm quiet}$ varies from 1
%to more than 10 Gyr. For smaller masses, the feedback is too weak to
%yield a first burst, and the system quickly reaches a self-regulated
%quasi-equilibrium level.  For much higher masses the feedback blows
%the ISM far out of the central regions where the gas is too dilute to
%dissipate substantially within a Hubble time.

%   Also more than one quiescence phase is possible. This is controlled by
%the feedback term $f$ in the stellar birth function $\Psi_b$: When the
%feedback parameter $T_s$ is reduced, the feedback factor $f$ is
%smaller for the same ''temperature'' and radius of the system, i.e.\ 
%less star formation takes place. Hence, the collapse will be deeper
%and the final reexpulsion will be stronger. As a result the system
%expands to large radii again, by this starting another single or small
%series of quiescence phases before reaching equilibrium.

\acknowledgements %%% Text of acknowledgements runs on after this command.
CT gratefully acknowledges support by the Univ.\ of Vienna and
by the DFG Priority Programme 1177 `Galaxy Evolution'.

%%% THE BIBLIOGRAPHY
%%%
%%% CONSULT SECTION 3 OF "INSTRUCTIONS FOR AUTHORS" FOR HOW TO USE NATBIB.
%%% AUTHORS ARE ENCOURAGED TO USE EITHER THE "THEBIBLIOGRAPY" ENVIRONMENT
%%% BY UNCOMMENTING (DELETING THE "%" SYMBOL) THE COMMANDS BELOW, OR BY
%%% USING THE BIBTEX ENVIRONMENT. TO FIND OUT WHICH IS APPLICABLE TO YOUR
%%% CONTRIBUTION, CONSULT THE VOLUME EDITORS FOR YOUR PROCEEDINGS.
%%%

\end{document}